# A Blockchain-Based Architecture for Traffic Signal Control Systems

Wanxin Li [a], Mark Nejad [a], Rui Zhang [b]

[a] Department of Civil and Environmental Engineering
[b] Department of Computer and Information Sciences
University of Delaware
Newark, DE 19716, United States
{wanxinli, nejad, ruizhang}@udel.edu

*Abstract*—Ever-growing incorporation of connected vehicle (CV) technologies into intelligent traffic signal control systems brings about significant data security issues in the connected vehicular networks. This paper presents a novel decentralized and secure by design architecture for connected vehicle data security, which is based on the emerging blockchain paradigm. In a simulation study, we applied this architecture to defend the Intelligent Traffic Signal System (I-SIG), a USDOT approved CV pilot program, against congestion attacks. The results show the performance of the proposed architecture for the traffic signal control system.

*Keywords-blockchain; connected and automated vehicles; data security; data credibility; internet of things; internet of vehicles; vehicular networks; hyperledger; traffic signal control*

## I. INTRODUCTION

Emerging adaptive traffic signal control systems incorporate real-time traffic data in their signal phase and timing (SPaT) mechanisms to improve the performance of intersections (e.g., safety and throughput). However, centralized traffic signal control systems and their datacenters can be attacked by receiving and processing malicious messages from connected vehicles in the traffic network. These malicious messages can include false information about vehicle IDs, locations, trajectories, etc. Systematic malicious attacks are a major challenge for traffic datacenters that need to validate a large amount of vehicular data for making decisions in real time. Without a trustable defending mechanism, malicious information could lead to serious consequences in a traffic network such as collisions [1] and congestions [2]. In this paper, we present a blockchain-based architecture to defend intelligent traffic signal control systems against information and data attacks by transforming the conventional connected vehicle network into a trustable and transparent decentralized network.

As an emerging computer network technology, blockchain was first invented in a cryptocurrency system, Bitcoin [3]. In the past few years, blockchain-based system designs have come a long way, and they have been successful in various decentralized applications [4, 5]. The nature of traceability and transparency in blockchain has a suitable match with increasing demands for data security in the connected-vehicle networks. However, most blockchain-based applications depend largely on digital tokens for the system design. This limits blockchain technology to be implemented mostly in cryptocurrency related systems. In this paper, we extend blockchain technology from classic cryptocurrency systems into traffic signal control systems. Blockchain not only links vehicles and infrastructures together in a decentralized network but also it works as a distributed and immutable ledger to automatically record vehicular information with timestamps. Furthermore, this distributed ledger provides trustable input data directly for intelligent traffic signal control systems.

### A. Our Contributions

We address the problem of data security in CV-based traffic signal control systems. These intelligent systems receive and process a certain number of arrival vehicle information as input table to generate optimal traffic signal plans at each intersection. Due to limited computational power in real-time processing and their centralized algorithms and datacenters, they are vulnerable if the input table contains spoofing vehicle information. To defend CV-based traffic signal control systems against malicious data attacks, we designed a blockchain-based decentralized architecture.

To the best of our knowledge, this is the first study exploring the blockchain paradigm in CV-based traffic signal control systems. Our proposed architecture introduces i) a customized blockchain network for connected vehicles; ii) and a consensus protocol design for validating source data. For the blockchain network, we choose Hyperledger Fabric [6] framework as the developing platform. Comparing with other blockchain frameworks, Hyperledger Fabric provides more flexibility for non-cryptocurrency system design.

In this study, we developed a blockchain prototype network. In addition, we perform simulations that show our prototype network can maintain a trustable distributed ledger for recording arrival vehicle information. For the consensus protocol, we designed a new mechanism to avoid attacker sending spoofing source information to the blockchain network. We add Roadside Units (RSU) and witness vehicles together as references for other nodes in the network to validate every piece of vehicle information before recording permanently in the blockchain network.

To show how our proposed architecture contributes to a realistic CV-based traffic signal control system, we applied our architecture to defend the vulnerable USDOT Intelligent Traffic Control System (I-SIG) [7] in a case analysis. In our architecture design, we utilize the distributed ledger on blockchain networks as input for traffic signal controller, which will avoid spoofing attack to the original datacenter.

*B. Organization*

The rest of the paper is organized as follows. In Section II, we present previous research in CV network attacking and recent progresses in blockchain applications. In Section III, we describe a full architecture design for the vehicular network transform, a blockchain framework preliminary, and we present our blockchain-based network, consensus protocol, and the workflow process. To further illustrate how our blockchain based architecture works to defend realistic intelligent traffic signal control systems, we choose Intelligent Signal Control System (I-SIG) [7] as a case analysis in Section IV. In Section V, we performed extensive experiments to test the robustness and performance of our developed CV blockchain network. In Section VI, we analyze the security of the proposed architecture against potential attacks. In Section VII, we conclude this study and present directions for future research.

## II. RELATED WORK

*A. Data Spoofing Attack in CV Networks*

Similar to many kinds of intelligent traffic signal control systems, I-SIG system [7] take arrival vehicle information as input table and generate optimal signal plans at intersection. In a recent work, Chen et al. [2] showed that the I-SIG system is vulnerable in the signal control algorithm level. Due to limited computation power, the signal controller cannot handle data validation in the real-time processing requirement, usually 5-7 seconds. They conducted the V2I attacking strategy by spoofing one vehicle information in the arrival table which caused congestion. Previously, Amoozadeh et al. [1] presented that spoofing attack in a V2V-based network can cause significant instability and even collisions. In another work, Dominic et al. [8] reported new attack surfaces and data flow in V2V-based network. Note that V2I attacks can affect all vehicles in the same network as I-SIG attacking scenario [2] whereas V2V attacks that can affect a certain group of vehicles.

*B. Blockchain Technology in Transportation*

In recent years, exploring the Blockchain paradigm in general transportation field has attracted a great deal of attention (e.g. [9-11]). Founded in August 2017, Blockchain in Transport Alliance (BiTA) has attracted more than 450 members around the world and became the largest commercial blockchain alliance [12]. These members are primarily from freight, logistics, technology companies and also academic institutes. The mainstream for implementing blockchain technology in transportation industry are freight tracking and food supply chain management. For instance, IBM has been working with retail giant Walmart to develop an efficient blockchain-based tracking system for food supply chain [13]. The blockchain technology helps Walmart to reduce tracing product time from weeks to seconds. This gives the company the ability to not only track where the food came from quickly but also how it was processed and distributed safely and responsibly.

Some studies have presented the possibility of implementing blockchain technology in forensic investigation. A recent study proposed a forensic investigation framework for IoT using blockchain, which is called FIF-IoT [14]. In addition, Guo et al. [15] proposed a blockchain-inspired "proof of event" mechanism for accident recording system in CAV network. Compared to these studies, our work focuses on blockchain-based system design in a new field that improves data security for CV-based traffic signal control systems.

## III. ACHITECTURE DESIGN

*A. Vehicular Network Transform*

In a conventional centralized CV network (Fig. 1), every traffic signal control system has to set up its own datacenter that runs all the codes and receives all the data. In addition, vehicles interacting with this control system must communicates with its centralized datacenter. Due to lower transparency and the single point of failure, a centralized architecture is not suited for creating trustable connected vehicle networks that have frequent real-time data transmissions.

We propose a blockchain traffic data network (Fig. 2) in which decentralization brings vehicles closer. Instead of having a central server and a database, the blockchain is a network and a database all in one [16]. It creates a vehicle-to-vehicle and vehicle-to-infrastructure network that share all the data. Any vehicle connected to the blockchain talks to all the other vehicles and infrastructures in the network. Thus, there are no more centralized server but only connected vehicles and infrastructures that reach into agreements on the network.

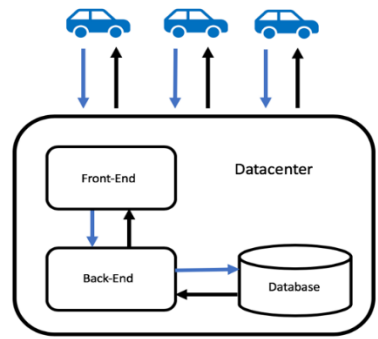

Figure 1. Central Server Vehicle Network

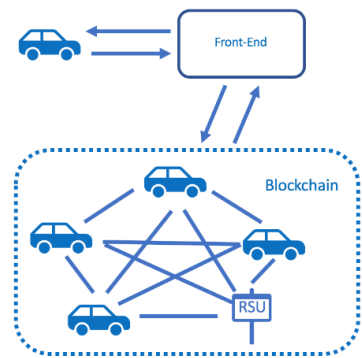

Figure 2. Blockchain-Based Vehicle Network

## B. Blockchain Framework Preliminary

In our architecture design, we choose Hyperledger Fabric [6] as the developing platform. It is the common platform for various mainstream blockchain systems. Comparing with older frameworks like Bitcoin [3], both Hyperledger Fabric and Ethereum [17] can provide programmable portion which is called Smart Contract [18]. Smart Contract is where the business logic of a blockchain network runs. We choose Hyperledger Fabric [6] instead of Ethereum [17] because the former provides more flexibility and modularity for blockchain implementation among cross-industries [19]. Most popular frameworks like Ethereum [17] cannot avoid digital tokens in system design. This restricts blockchain technology to serve well only in cryptocurrency related system. In addition, Hyperledger Fabric [6] has a cost-effective approach towards transactions since no mining process from a cryptocurrency design is needed anymore. On the contrary, both Bitcoin [3] and Ethereum [17] require nodes to mine transactions by longer processing time and significant consumption of computation hardware and electricity.

In a connected vehicular network, we utilize blockchain technology as a distributed ledger that records every vehicle information including VIN, Location (GPS) and trajectory in ledger (Fig. 3). In addition, Blockchain technology automatically add timestamp for each record, which makes it traceable. For this purpose, we don't involve digital tokens in architecture design level to avoid adding unnecessary components and overheads. On the other hand, the flexibility and modularity in Hyperledger Fabric have been proved well in freight tracking and food supply chain systems like the IBM and Walmart project [13]. These precedents give us an appropriate launchpad for leveraging blockchain technology into connected vehicular network. Instead of recording vehicular information in a vulnerable and centralized server, blockchain technology creates a transparent and trustful decentralized database providing reliable information to the traffic signal control systems.

As Figure 4 [20] shows, Hyperledger Fabric [6] is a highly modularized framework for developing full-stack blockchain networks. We first describe a blockchain network in four programmable parts: Model File, Script File, Access Control and Query File. Model File is where we define all the objects in the network. All the response functions are written in Script File. Hyperledger Fabric also provides Access Control to restrict data access to certain roles in the network. As for Query File, it works similar with conventional database query definitions. Except for the Model File, the other three parts are pluggable according to the application requirements. Then, we package up these files into one Business Network Archive file and deploy it into a running blockchain network. This blockchain network can be accessed and tested in a front-end webpage.

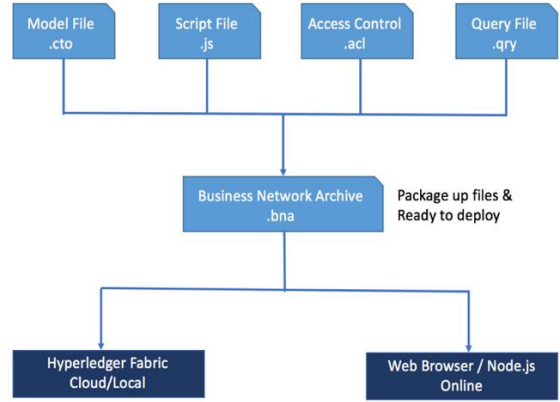

Figure 3. Distributed Ledger on Blockchain

Figure 4. Hyperledger Fabric Infrastructure

## C. Developing the Blockchain Network

We developed a blockchain network prototype based on Hyperledger Fabric framework. We identified each vehicle by its VIN number. Our blockchain network maintains a distributed ledger for sharing and recording of arrival vehicle information as input for the traffic signal control systems. As shown in Figure 3, we define arrival vehicle information in the Model File as follows:

| Define Arrival Vehicle Information |
|---|
| 1.   Vehicle_Info { |
| 2.      Record_ID |
| 3.      VIN |
| 4.      GPS{ |
| 5.         Longtitue |
| 6.         Latitude |
| 7.      } |
| 8.      Trajectory{ |
| 9.         Speed |
| 10.        Accelartion |
| 11.     } |
| 12.     Timestamp |
| 13. } |

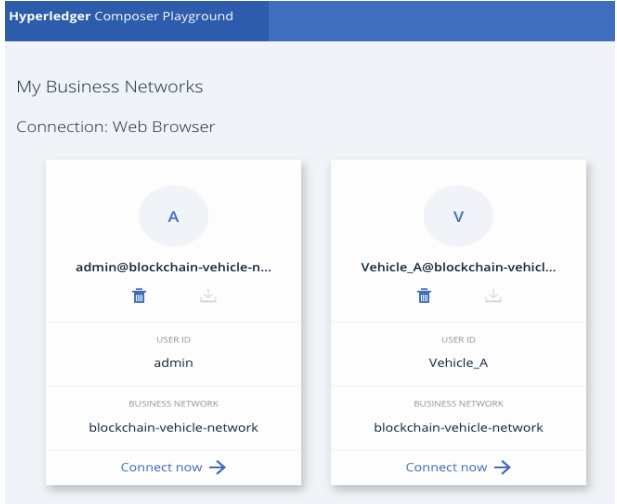

Figure 5. Blockchain Network Webpage UI

In order to make the ledger immutable, we grant Access Control rule for all participants. Each participant (i.e. vehicle, RSU, and traffic signal controller) only have ADD or READ operation access for ledger records. Therefore, no one can modify data in the ledger. We use Hyperledger Composer Tool [21] to generate the deployable unit file (.bna) and deploy it on the blockchain network. Hyperledger Composer Tool [21] also provides a webpage interface for connecting and testing the blockchain network (Fig 5). Each participant has an ID registry for connecting to the blockchain network, and we assign the traffic signal controller as the administrator. Other users' roles are either a vehicle or an RSU.

### D. Consensus Protocol Design

By deploying blockchain technology into a connected vehicle network, we can guarantee data immutability and traceability in a decentralized ledger. For this purpose, we design a consensus protocol for the network to validate the source vehicle information. After validation, our blockchain network records vehicle information permanently. Classic blockchain protocol in cryptocurrency can validate new transactions by checking hash code of tokens and the previous transaction history [22]. This is trustable since all tokens were carefully defined and encrypted as source data within the system from beginning. However, we do not involve digital tokens concept into the proposed connected vehicular network. Consequently, the consensus protocol needs a creative design.

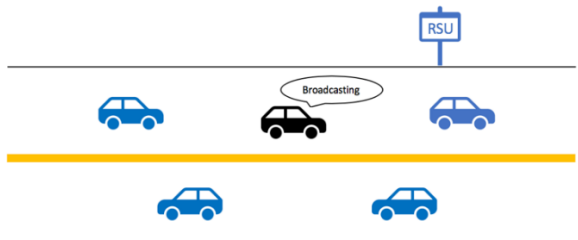

Figure 6. Broadcasting Scenario

Vehicles broadcast their information among the blockchain network. For consensus protocol design, we add Roadside Units (RSU) as nodes into our blockchain network. Then, we introduce witness vehicles and nearby RSU together as references for validating source information. In this scenario (Fig. 6), if a broadcasting vehicular information is matched with references from its nearby RSUs and witness vehicles, the source vehicular information is trustable and we let the blockchain network to record it. On the contrary, if the source vehicular information cannot match with the references, we treat this as a malicious vehicular information and will not let blockchain network to record it. At the same time, we can locate and add this vehicle as an attacker into a blacklist. The consensus algorithm is represented as the following pseudo-code:

Consensus Algorithm

```
1.   s = source data;
2.   r = reference data;
3.
4.   Function validation (s, r) {
5.       l = distributed ledger;
6.       b = blacklist for recording attacker;
7.       if (b.find(s) == true ) {
8.           reject;
9.       } else {
10.          if (s == r) {
11.              l.add(s);
12.          } else {
13.              reject;
14.              b.add(s);
15.          }
16.      }
17.  }
```

### E. Workflow Process

Combining the above-mentioned four parts, we reach at the full view of our blockchain-based architecture for connected vehicular networks. As shown in the flowchart (Fig. 7), blockchain technology makes vehicular information transparent and trustable by providing protocol and cryptography on a decentralized network. When a connected vehicle broadcasts its information, the other nodes in the same network first validate this information by comparing it with references from nearby RSUs and witness vehicles. If the source information is false, blockchain network will not record this piece of information for the traffic control processes and record the malicious attack and the attacker. If the source information is correct, blockchain network will record and share it on a decentralized ledger. Blockchain technology automatically calculates each vehicular information into a hash code. Since every node including connected vehicles and RSUs saves all the data in the network, a spoofing attack can be quickly found by a peer-to-peer check. All the nodes will reach into an agreement for checking data and this process can be finished in real-time, within milliseconds.

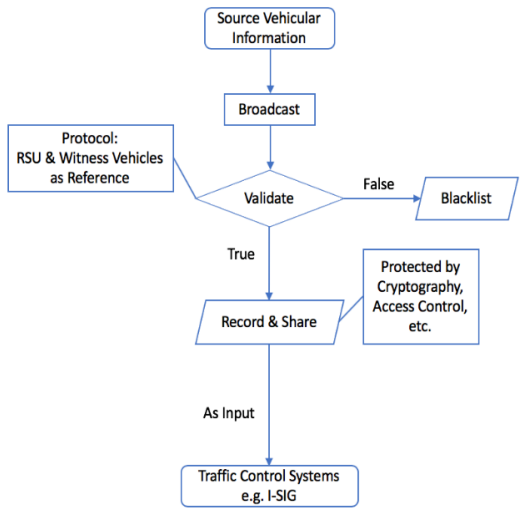

Figure 7. Architecture Flowchart

## IV. CASE ANALYSIS

To show how our decentralized architecture works in defending traffic signal control systems, we employ the I-SIG system [7] as a case analysis, which is an intelligent signal control system for connected vehicles. As one of USDOT approved CV Pilot Programs, this system has been deployed in New York City, Tampa and Wyoming since 2016 [7].

The I-SIG system takes arrival vehicles' BSM (Basic Safety Message) messages, which contain locations and trajectories, as an input table to calculate and generate signal plans at each intersection (Fig. 8).

In a recent paper [2], Chen et al. showed that the I-SIG system can be easily attacked in order to create congestions (Fig. 9). In their work, they first showed that I-SIG [7] is not able to validate arrival vehicles' data in real-time. They modified one vehicle's location and trajectory data in the arrival table. As a result, this straight forward attacking strategy can account for a blocking effect that jams the whole intersection.

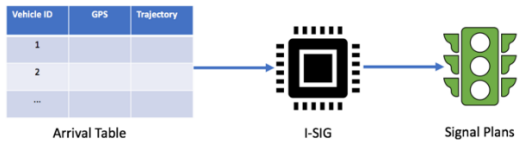

Figure 8. Original I-SIG system

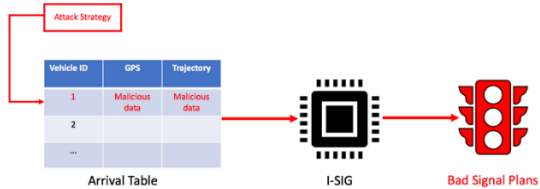

Figure 9. Attacking I-SIG system

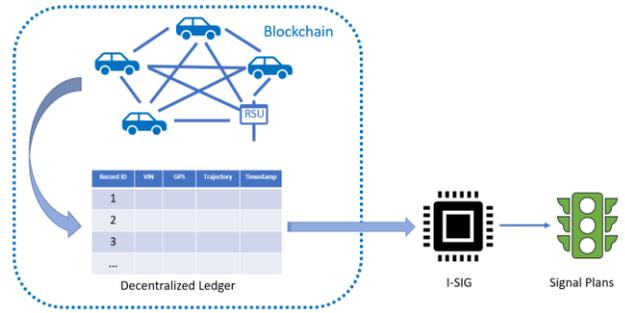

Figure 10. I-SIG system with Blockchain Technology

This kind of attacking strategies can work successfully in a traffic signal control system relying on centralized vehicular networks. Our defending strategy is to leverage our blockchain-based architecture to transform the original centralized vehicular network in a decentralized one (Fig. 10). Instead of receiving and saving all vehicular information in a vulnerable datacenter, we record and share the information in a transparent and trustable decentralized ledger with traceable timestamps on the blockchain network. We show that blockchain can keep data immutable due to its decentralized cryptographic mechanism. We also introduce consensus protocol that combines nearby RSUs and witness vehicles as s for the source vehicular information validation. In this way, the decentralized ledger provides clean data input for traffic signal control systems such as I-SIG [7]. If an attacker is trying to modify the record on the blockchain, the network can quickly locate and reject the attack.

## V. EXPERIMENTS

### A. Experimental Setup

We conducted simulations to test the robustness and performance of our blockchain framework under spoofing attacks. We deployed the blockchain network on Hyperledger Composer [21], which will maintain a distributed ledger for recording and sharing arrival vehicle information that contains VIN, GPS, trajectory and timestamp. We simulated sending and recording arrival vehicle information process by initializing 20 records on the distributed ledger (Fig. 11). Based on our consensus protocol design, the initialized records on the distributed ledger are validated arrival vehicle information. We access the blockchain and conduct experiments on macOS High Sierra operating system with 2.9 GHz Intel i5 processor with 60 Mbps bandwidth Wi-Fi connection as the default settings. We simulated the attacking strategy by trying to modify records on the ledger. We then checked the response of our blockchain framework against attacks and record its performance via Chrome DevTools. To provide more insights for hardware and internet requirements of our proposed architecture in real CV environment, we conducted a series of experiments with different participant numbers, network speed, and processor speed.

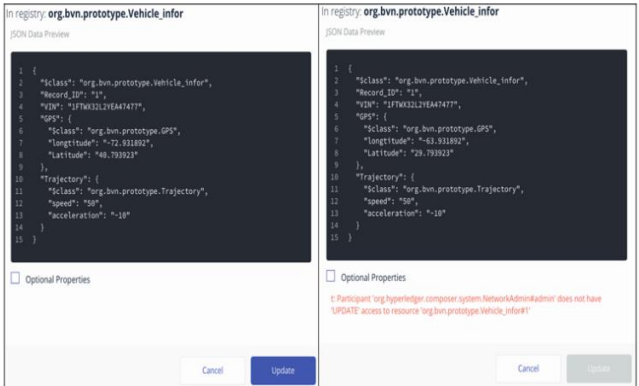

Figure 11. Initializing Arrival Vehicle Information

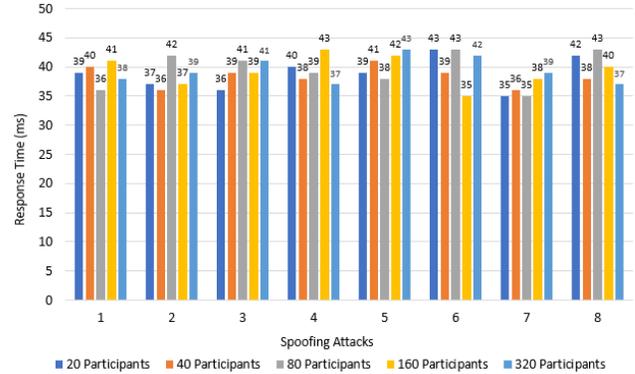

Figure 13. Response Time When Changing Participant Number

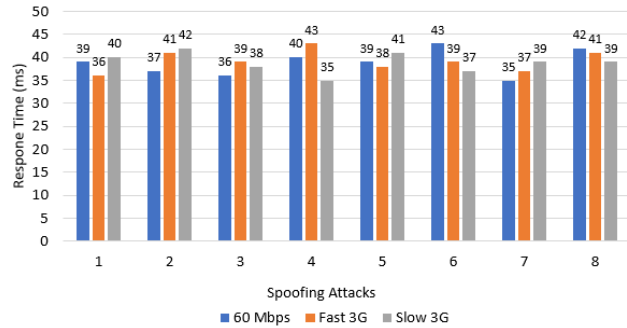

Figure 14. Response Time When Changing Network Speed

Figure 12. Response Against Modifying Record

## B. Response Against Attack

Protected by blockchain technology, our prototype framework will reject vehicular spoofing attacks 100% of the time successfully in real-time. Once an arrival vehicle information is saved on the distributed ledger, it does not allow any participant to modify it. Our blockchain framework rejects and pops out a warning message for any attempt to modify the records (Fig. 12). We use Chrome DevTools to record the performance and found that the average response time is on average 39 ms on the default hardware and internet settings. Considering that intelligent traffic control systems such as I-SIG take 5 to 7 seconds for processing signal plans, our proposed architecture will easily meet requirements for real-time operation and protect vulnerable traffic control systems.

## C. Change in the Participant Number and Network Speed

In a blockchain network, every participant runs the same code and saves the same data in a distributed way. Theoretically, our framework performance cannot be affected by participant number or network speed when attack happens.

To change the number of participants, we increase original arrival vehicle records in ledger from 20 to 40, 80, 160 and 320 and conduct 8 attacks separately. The average response time against attacks keeps around 39 ms as shown in Fig. 13.

To change the network speed, we changed network settings from default 60 Mbps bandwidth Wi-Fi to fast 3G and slow 3G. Similarly, the average response time against 8 attacks is still around 39 ms (Fig. 14). In extreme condition, we set the attacker offline (i.e. it cannot access to the ledger even locally). However, the ledger will restore once network connects. Note that changing network speed will only affect performance of adding and sharing new data (arrival vehicle information in our case analysis) on the distributed ledger.

## D. Change in the Processor Speed

Since each participant runs code on its own processor and their hardware specifications are different, we tested our proposed architecture on different settings. We conduct this experiment by throttling default CPU speed slower. In CPU 4 times slower scenario, we get the average response time at 74 ms. In another setting, we slowdown the CPU 6 times. Loading webpage like popping out warning message becomes slower, in seconds. However, the back-end response process keeps at around 118ms. Figure 15 shows the response results based on 8 attacks in default CPU, 4 times slowdown and 6 times slowdown configurations. The results show that our framework will still work well in a CV environment with low-tier processors.

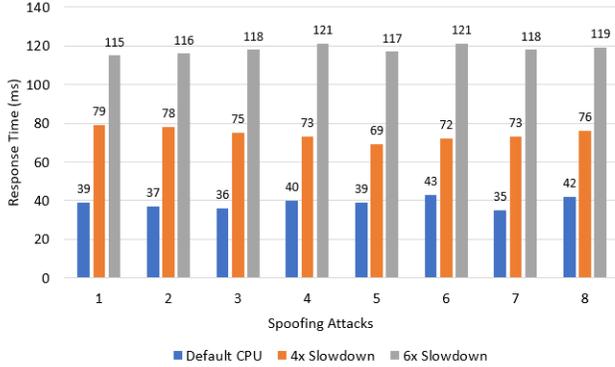

Figure 15. Response Time When Changing Processor Speed

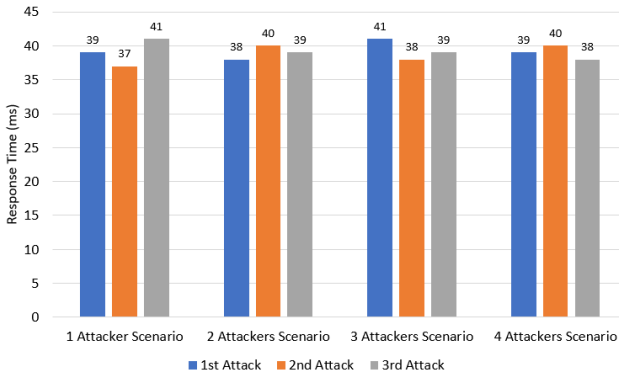

Figure 16. Default Computer Response Time Against Multiple Attacks

### E. Multiple Attacks at the Same time

We conduct above experiments based on one signal attack scenario [2] in the I-SIG system [7]. In our last experiment, we test the response performance against multiple attacks at the same time. Based on the Part C results, our framework performance is not affected by other participants since the framework distributes codes and data on each participant's hardware. Therefore, multiple attacks do not affect the response performance. To verify this conjecture, we deployed our framework on a Local Area Network (LAN) and add three more computers A, B and C into the network as potential attackers. Although these computers have different processors, we focus on observing response time on our default computer, which has a 2.9 GHz i5 processor. To find the relationship between response time and multiple attack numbers, we conducted this experiment in four rounds: (1) default computer is the only attacker; (2) default computer and computer A are attackers; (3) default computer, computer A and computer B are attackers; (4) All four computers are attackers. We repeat 3 times in each round and record response performance for the default computer in figure 16. The results show that when there are multiple attacks at the same time, the framework can reject the attacks and keep response time of 39 ms for the default computer.

Our framework's performance is not affected by the misreports of multiple participants. Our Blockchain-based framework can transform the conventional connected vehicular network into a decentralized one in which not only the data but also the codes are saved and executed on each participant's hardware.

## VI. SECURITY ANALYSIS

In this section, we will analyze the security of the proposed blockchain-based decentralized architecture for the connected vehicular networks.

### A. Spoofing Source Vehicle Information

In a connected vehicle network, it is possible for on-site attackers to broadcast spoofing source vehicle information, such as false locations or trajectories. In order to avoid this kind of attack, we first add RSUs as nodes in our blockchain network. We then combine nearby RSUs and witness vehicles as references for consensus protocol. By adding this consensus protocol into our architecture, all participants in the same network will achieve agreement on validating the source information process. If the source information is matched with reference information from RSUs and witness vehicles, the blockchain network approves and saves it permanently on the distributed ledger. If the attacker is broadcasting spoofing vehicle information such as false location or trajectory information, this information does not match with reference information from RSUs and the witness vehicles. Our architecture rejects this spoofing information and adds the attacker into a blacklist.

### B. Recorded Data Attack

Blockchain technology keeps data immutable. It ensures data security by saving data in a distributed ledger, peer-to-peer check and various types of pluggable cryptographic algorithms including hash digest [23] and Merkle tree [24]. As mentioned in Section III Part B, Hyperledger Fabric [6] also provides Access Control to restrict data access to certain users in the network. The Access Control is implemented in a way that the participants can only read or add new data in the distributed ledger, but they cannot make any modifications. When an attacker is trying to modify the ledger record, our blockchain framework rejects and pops out a warning message immediately.

### C. Multiple Attacks at the Same Time

We extended I-SIG attacking strategy presented in [2] from single attack to multiple and simultaneous attacks. The proposed architecture rejects all the attacks and keeps response performance the same for each participant. This shows that blockchain technology can fully transform connected vehicle network into a decentralized architecture. There is no centralized server in the network and each participant runs the code on its own hardware.

### D. Majority Attack

A majority attack or 51% attack is an extreme attacking scenario when there is a super node that tries to manipulate the blockchain network, which has more computational power than the rest of the nodes. This only exists theoretically in mining-based blockchain frameworks such as Bitcoin and Ethereum. In our proposed architecture, the blockchain network maintains a distributed ledger for recording arrival vehicle information. Our architecture is resilient to majority attack since we avoided redundant digital tokens, transactions and mining process by employing the flexible Hyperledger Fabric [6] framework.

## VII. CONCLUSION

In this paper, we designed a blockchain-based and decentralized architecture for connected vehicular networks. Targeting a promising blockchain implementation in a new area, we refine the workflow process in our vehicular network representation. In addition, we developed a blockchain prototype network and consensus protocol. To show how our architecture works in a realistic traffic signal control system, we used I-SIG system [7], which is under USDOT approved CV Pilot Program, as a case analysis. By transforming the original centralized vehicular network into a decentralized one, we defend the original vulnerable I-SIG system [7] against malicious attacks. In addition, we conducted a series of simulations to analyze the response performance under different settings.

This study serves as the first step for migrating blockchain technology from cryptocurrency systems into traffic signal control systems. Future research directions include: (1) novel consensus protocol designs for validating broadcasted source vehicular data when a systematic group attack happens on site, when both nearby RSUs and witness vehicles cooperate with the attacker to send spoofing reference; (2) other realistic intelligent traffic control systems based on connected vehicles; (3) flexible blockchain framework developments for cross-industries implementations.